\documentclass[twocolumn,showpacs,aps,showkeys]{revtex4}
\usepackage{graphicx}
\usepackage{amsmath}

\begin{document}

\title{The origin of diffusion: the case of non chaotic systems}

\author{Fabio Cecconi}
\email[corresponding author:]{fabio.cecconi@roma1.infn.it}
\affiliation{
Dipartimento di Fisica, Universit\`a di Roma "La Sapienza",
P.le A.Moro 2, I-00185 Rome and INFM Center for Statistical Mechanics and
Complexity, Italy}

\author{Diego del-Castillo-Negrete}
\affiliation{Oak Ridge National Laboratory, MS 8071, Oak Ridge, TN 37831-8071}

\author{Massimo Falcioni}
\affiliation{
Dipartimento di Fisica, Universit\`a di Roma "La Sapienza",
P.le A.Moro 2, I-00185 Rome and INFM Center for Statistical Mechanics and
Complexity, Italy}

\author{Angelo Vulpiani}
\affiliation{
Dipartimento di Fisica, Universit\`a di Roma "La Sapienza",
P.le A.Moro 2, I-00185 Rome and INFM Center for Statistical Mechanics and
Complexity, Italy}
\date{\today}

\begin{abstract}
We investigate the origin of diffusion in non-chaotic systems.  As
an example, we consider $1$-$d$ map models whose slope is everywhere
$1$ (therefore the Lyapunov exponent is zero) but with random quenched
discontinuities and quasi-periodic forcing. The models are
constructed as non-chaotic approximations of chaotic maps showing
deterministic diffusion, and represent one-dimensional versions of a
Lorentz gas with polygonal obstacles (e.g., the Ehrenfest wind tree
model). In particular, a simple construction shows that these maps
define non-chaotic billiards in space-time.  The models exhibit, in a
wide range of the parameters, the same diffusive behavior of the
corresponding chaotic versions. We present evidence of two sufficient
ingredients for diffusive behavior in one-dimensional, non-chaotic
systems: i) a finite-size, algebraic instability mechanism, and
ii) a mechanism that suppresses periodic orbits.
\end{abstract}



\maketitle

\section{Introduction}
The presence of randomness in many natural processes is traditionally
ascribed to the lack of perfect control on observations and
experiments.  Then, usually,``disorder'' and ``irregular behaviors''
are considered only as unavoidable factors which, in phenomena
modellizations, can be described effectively through external
sources (random numbers in computer simulations, noise in stochastic
equations etc.). However randomness may also have another origin
associated to local instabilities of the deterministic
microscopic dynamics of a system, i.e. microscopic chaos (MC). For instance,
macroscopic diffusion of a particle in a fluid can be interpreted as a
consequence of chaos on microscopic scales generated by collisions of
the particle with the surrounding others.

The possibility of
describing the Brownian behavior in terms of deterministic motion has
stimulated a large amount of work in the field of deterministic
diffusion to study purely deterministic systems exhibiting
asymptotically a linear growth of the mean square
displacement~\cite{ChaoDiff,ChaoDiff1,KD}. The deterministic
interpretation of transport phenomena establishes a close connection
between transport and chaos theory, and, at least in principle,
macroscopic properties such as transport coefficients (viscosity,
thermal and electrical conductivity, diffusion etc.) could be directly
computed through indicators of chaos \cite{GasNik,Morriss,Dorfman} or
by applying the standard techniques of chaos theory such as periodic
orbit expansion \cite{Artuso,Vance,CEG95,Gaspard}. However,
considering randomness and transport in general as signatures of
MC raises subtle and fundamental issues in statistical
physics that have to be carefully discussed an investigated.  First,
there is the main problem of conceiving and realizing experiments able
to prove, beyond any doubt, the existence of MC in real systems. An
experiment in this direction was attempted in ref.~\cite{Nature} but
the conclusions were object of criticism by many authors, as we
discuss later. Besides, for real systems, the definition of MC appears 
to be ambiguous. Indeed, real systems involve a large number of
degrees of freedom and their model description entails taking 
the thermodynamic limit but, in this limit,  
the concept of Lyapunov exponent becomes metric dependent so not well 
established~\cite{GS}. Moreover, the large number of degrees of freedom poses
severe limitations also in the applicability of standard
techniques to distinguish between chaotic and stochastic
signals~\cite{GP,CP,GW} when data are extracted from real experiments.
In fact, the noise-chaos distinction requires a extremely fine
resolutions in the observations that is impossible to reach when one
deals with systems with large dimensionality \cite{DCH,DC,Cencio}.

References~\cite{Nature,Experiment} summarized an ingenious experiment
devised to infer the presence of chaos on microscopic scales.  In the
experiment the position of a Brownian particle in a fluid was recorded
at regular time intervals. The time series were processed by using the
same techniques developed for detecting chaos from data
analysis~\cite{CP,GW}. Authors measured a positive lower bound for
the Kolmogorov-Sinai entropy, claiming that this was an experimental
evidence for the presence of microscopic chaos. However in the papers 
~\cite{DCH} and ~\cite{DC} it was argued that a similar result holds for 
a non-chaotic system too, hence the experiment cannot provide a conclusive 
evidence neither for the existence of microscopic chaos nor for the relevance
of chaotic behavior in diffusion phenomena. 
Such papers shown that, through the analysis performed in 
Refs.~\cite{Nature,Experiment}, there is no chance to observe differences 
in the diffusive behavior
between a genuine deterministic chaotic systems as the 2-d Lorentz gas
with circular obstacles~\cite{Dorfman} and its non chaotic variant:
the wind-tree Ehrenfest model. The Ehrenfest model consists of free
moving independent particles (wind) that scatter against square
obstacles (trees) randomly distributed in the plane but with fixed
orientation. Due to collisions, particles undergo diffusion, however
their motion cannot be chaotic because a reflection by the flat planes
of obstacles does not produce exponential trajectory separation; the
divergence is at most algebraic leading to zero Lyapunov exponent.
Such considerations can be extended to squares obstacles with random
orientations and to every polygonal scatterers, so there is a whole
class of models where diffusion occurs in the absence of chaotic
motion. A difference between the 2-$d$ chaotic and 
non-chaotic Lorentz gas relies on the presence of periodic 
orbits~\cite{DC}. 

The question that naturally arises ``What is the microscopic
origin of the diffusive behavior?''  remains still open.  In this
paper, focusing on this problem, we extend the analysis of Dettman and
Cohen showing that standard diffusion may also occur in
one-dimensional models where every chaotic effect is surely absent.
These systems, being one-dimensional, can be much easily studied than
wind-tree models without spoiling the essence of the problem. Accurate
results are obtained with relatively small computational efforts, and
the interpretation of the results is clearer since we are able to
better control and quantify the effects of the local instabilities.
We shall see that, in such models, the relevant ingredients to obtain
diffusion in the absence of deterministic chaos is the combined action
of a ``finite size instability'', quenched disorder and a
quasi-periodic perturbation.  Where by ``finite size instability'' we
mean that infinitesimal perturbations are stable, while perturbations
of finite size can grow algebraically. The interplay between these three
features guarantees the system to be still diffusive as it would be 
chaotic.
This outcome is another striking effect of finite size instabilities
which make the behaviors of certain non-chaotic system similar, to
some extent, to those generated by genuine chaotic systems
(\cite{PLOK,Cecconi,Torcio,LRB}). 
In ref.~\cite{zasl} the reader can find a
study of an interesting system with weak mixing properties and zero
Lyapunov exponent, i.e. an intermediate case between integrability and
strong mixing. 

The paper is organized as follows. In section II, we first recall the
general features of $1$-$d$ models showing deterministic diffusion.
Then we present two non-chaotic, one-dimensional models that exhibit
deterministic diffusion when properly perturbed.  In section III we
characterize numerically the diffusion properties of the models and
their dependence on the external perturbation.  Section IV shows
how these models can be viewed as non-chaotic space-time billiards,
and discusses in more detail the role of periodic orbits.  Section V
is devoted to conclusions and remarks.

\section{Non-Chaotic Models for Deterministic Diffusion}
Motivated by the problem of deterministic diffusion in non-chaotic
Lorentz systems (i.e. with polygonal obstacles) we introduce models
that, somehow, represent their one-dimensional analog.  We begin
considering one of the simplest chaotic model that generates
deterministic diffusion: a $1$-$d$ discrete-time dynamical system
on the real axis
\begin{equation}
x^{t+1} = [x^t] + F(x^t-[x^t]) \,  ,
\label{eq:chaos}
\end{equation}
where $x^t$ is the variable performing the diffusion (position of a
point-like particle) and $[\ldots]$ denotes the integer part.
$F(u)$ is a map defined on the interval $[0,1]$
that fulfills the following properties
\begin{description}
\item[i)] $|F'(u)|>1$, so the system has a positive Lyapunov
exponent.
\item[ii)] $F(u)$ must be larger than $1$ and smaller than $0$ for
some values of $u$, so there exists a non vanishing probability to escape
from each unit cell (a unit cell of real axis is
every interval $C_{\ell} \equiv [\ell,\ell+1]$, with $\ell \in {\bf Z}$).
\item[iii)] $F_r(u)=1-F_l(1-u)$, where $F_l$ and $F_r$ define the map
in $u\in [0,1/2[$ and $u\in [1/2,1]$ respectively. This anti-symmetry
condition with respect to $u= 1/2$ is introduced to prevent the
presence of a net drift.
\end{description}
The map, $u^{t+1}=F(u^t)$ (mod $1$) is assumed to be also ergodic.
One simple choice of $F$ is
\begin{eqnarray}
F(u) =
\left\{
\begin{array}{ll}
2(1+a) u   \qquad \qquad \quad \, \, \mbox{if~}  u \in [0,1/2[       \\
2(1+a) (u-1) + 1 \quad\mbox{if~}  u \in [1/2,1]
\end{array}\right.
\label{eq:chaos2}
\end{eqnarray}
where $a>0$ is the parameter controlling the instability.  This model
has been so widely studied, both analytically and numerically
\cite{KD,Klag,Klag1} that can be considered as a reference model for
deterministic diffusion. The deterministic chaos which governs the
evolution inside each unit cell is responsible of the irregular
jumping between cells and the overall dynamics appears diffusive, in
the sense that asymptotically the variance of $x^t$ scales linearly
with time. Following the literature on deterministic diffusion we
consider systems with no net drift ($i.e.$ $\langle x^t \rangle / t
\to 0 $ for $t \to \infty$). In this paper we will see that there is
no need to invoke chaos to induce diffusive behavior in 1-$d$
maps, as already shown in \cite{DC} in the context of 2-$d$
billiards.

The construction of the models begins by dividing the first half of
each unit cell $C_{\ell}$ into $N$ micro-intervals
$[\ell + \xi_{n-1},\ell + \xi_{n}[$ , $n=1,\dots, N$, with
$\xi_0=0 < \xi_1<\xi_2 < \dots <\xi_{N-1} < \xi_N=1/2$.
In each micro-interval the maps are defined by
prescribing a linear function $F_\Delta$ with unit slope. The map in the 
second half of the unit cell is determined by the anti-symmetry condition
iii).

In the first model $\{\xi_k\}_{k=1}^{N-1}$
is a uniformly distributed random sequence between $[0,1/2]$, so the size 
of the micro-intervals is {\em random} and
the function is defined as
\begin{eqnarray}
F_{\Delta}(u)=
\begin{array}{ll}
u - \xi_{n}   + F(\xi_n)  \qquad \quad \mbox{if~}
u \in [\xi_{n-1},\xi_{n}[ \;\; ,
\end{array}
\label{model_1}
\end{eqnarray}
where $F(\xi_n) $ is the chaotic map function in Eq.~(\ref{eq:chaos2})
evaluated at $\xi_n$.

\begin{figure}
\includegraphics[scale=0.35]{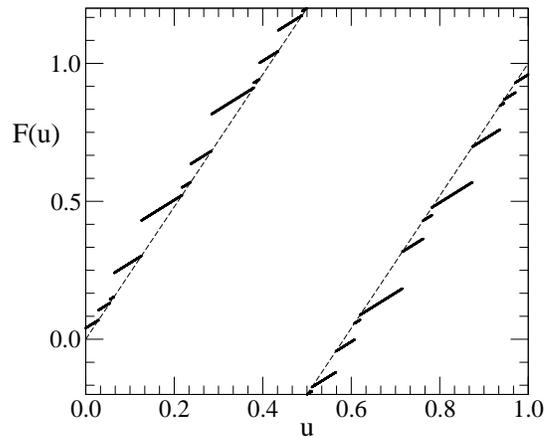}
\caption{\label{fig:map} Sketch of the random staircase map in
the unitary cell. The  parameter
$a$ defining the macroscopic slope is set to $0.23$.
Half domain $[0,1/2]$ is divided into $N=12$ micro-intervals of random size.
The map on $[1/2,1]$ is obtained by applying the antisymmetric
transformation
with respect to the center of the cell $(1/2,1/2)$}
\end{figure}

\begin{figure}
\includegraphics[scale=0.35]{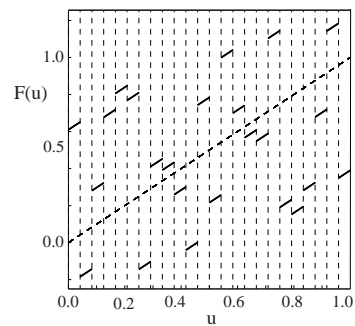}
\caption{\label{fig:model2} Sketch of model (\ref{model_2}) in the
unit cell.  In this case the half domain $[0,1/2]$ is divided into
$N=12$ uniform micro-intervals.  The map on $[1/2,1]$ is obtained by
applying the antisymmetric transformation with respect to the center of
the cell $(1/2,1/2)$.  In each micro-intervals the map is defined by
prescribing a linear function with unit slope and random height, with 
$\delta=0.2$ (see text).}
\end{figure}

In the second model, the size of the micro-intervals is {\em
uniform}, i.e. $|\xi_{n}-\xi_{n-1}|=1/N$, but the ``height" of the
function is {\em random}. That is,
\begin{eqnarray}
F_\Delta(u)=
\begin{array}{ll}
u - \tilde{\xi}_{n}   + \lambda_n  \qquad \quad \mbox{if~}
u \in [\xi_{n-1},\xi_{n}[ \;\; ,
\end{array}
\label{model_2}
\end{eqnarray}
where $\tilde{\xi}=(\xi_{n-1} + \xi_n)/2$, and the heigth at the mean point, 
$\lambda_n$, is a random variable distributed  uniformly 
in the interval $[-\delta,1+\delta]$, with $\delta \geq 0$.  
Figures~\ref{fig:map} and
\ref{fig:model2} show examples of these two models. Let us note that
the random variables $\xi_{n}$ and $\lambda_n$ are quenched variables,
i.e. in a given realization of a model they depend on the cell
$C_{\ell}$  (so one should properly write $\xi_{n}^{(\ell)}$ and
$\lambda_{n}^{(\ell)}$) but they do not depend on time.

As Fig.~\ref{fig:map} illustrates in model (\ref{model_1}) each cell
is partitioned in the same number $N$ of randomly chosen
micro-intervals of mean size $\Delta = 1/N$ and in each of them the
slope of the map (microscopic slope) is one.  For $\Delta\to 0$
(equivalently $N\to \infty$) we recover the chaotic
system~(\ref{eq:chaos}) but the limit $F_{\Delta}\to F$ has to be
carefully interpreted.  This kind of modification of the original
chaotic system is somehow equivalent to replacing circular by
polygonal obstacles in the Lorentz system \cite{DC}, since the steps with
unitary slope are the analogous of the flat boundaries of the
obstacles. The discontinuities of $F_{\Delta}$ produce a dispersion of
trajectories in a way similar to that of a vertex of polygons that
splits a narrow beam of particles hitting on it. However, at variance
with the $2$-$d$ wind-tree model, both randomness and forcing are
needed to attain a full diffusive behavior.  Note that, thanks to the
local preservation of the antisymmetry with respect to the cell
center, no net drift is expected and also the Sinai-Golosov
\cite{Sin,Gol,Quench} effect is absent.  
Also model (\ref{model_2}) is non-chaotic and the randomness is introduced
in the function but not in the partition of the unit cells.

Since $F_{\Delta}$ has slope $1$ almost everywhere, the chaoticity
condition i) is no more satisfied in these models.  However,
these models can exhibit diffusion provided
condition i) is replaced by the following
two requirements:
\begin{description}
\item[i-a)] presence of quenched
disorder,
\item[i-b)] presence of a quasi-periodic external perturbation.
\end{description}
In models (\ref{model_1}) and (\ref{model_2}), requirement i-a) is
satisfied by the randomness in the partition of the micro-intervals,
and the randomness in the linear function. To fulfill requirement i-b)
we introduce a time-dependent perturbation into Eq.~(\ref{eq:chaos})
\begin{equation}
x^{t+1} = [x^t] + F_{\Delta}(x^t-[x^t]) + \varepsilon \cos(\alpha t)
\label{eq:nochaos}
\end{equation}
where $\varepsilon$ tunes the strength of the external forcing,
$\alpha$ defines its frequency and the index $\Delta$ indicates a
specific quenched disorder realization. We underline again that the
linear piecewise map $F_{\Delta}$ explicitly depends on the cells
$C_{\ell}$ due to the disorder, which varies from cell to cell.
Even though by construction the system is not unstable in the Lyapunov
sense, it exhibits standard diffusion. In particular, numerical
simulations show a linear growth of the mean square displacement. It
is important to note that, for model~(\ref{model_1}) a uniform
partition of cell domains plus the quasi-periodic perturbation is not
sufficient to produce diffusive dynamics. Even in the presence of
quenched randomness (i.e. random partitions or random heights) alone,
the system spontaneously settles onto periodic or quasi-periodic
trajectories and does not diffuse.  The periodic and quasi-periodic
behavior is broken down, only when a time dependent perturbation is
explicitly superimposed onto the quenched randomness. Hence, we can
say that, in our models, the time plays somehow the role of the
missing spatial dimension with respect to the $2$-$d$ models of
ref.~\cite{DC}.  We will explore this idea in more detail in Sec.~V.

\section{Diffusion without chaos, numerical results.}
In this section, we characterize the diffusion properties of
model (\ref{model_1}), while in the next, we discuss diffusion on model
(\ref{model_2}).
 
\begin{figure}
\includegraphics[scale=0.35]{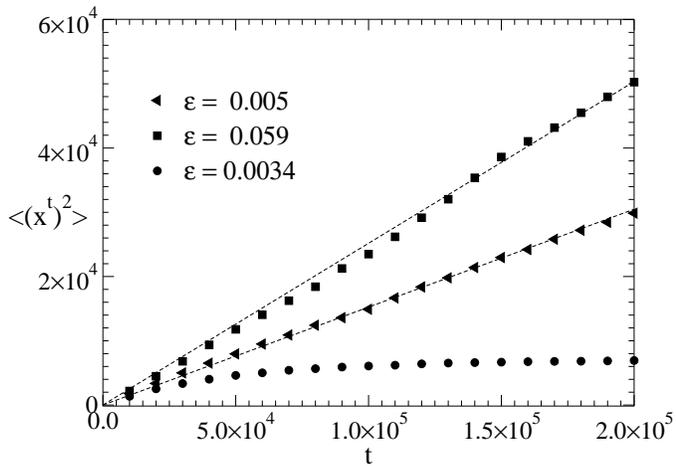}
\caption{Time behavior of $\langle (x^t)^2 \rangle$
for three different perturbation amplitudes $\varepsilon$, $a=0.23$ and
$N=100$ micro-intervals per cell.}
\label{fig:diffus}
\end{figure}

The diffusion coefficient,
\begin{equation}
D = \lim_{t\to\infty} \frac{1}{2t} \langle (x^t - x^0)^2 \rangle
\label{eq:diff}
\end{equation}
has been numerically computed from the slope of the linear asymptotic
behavior of the mean square displacement.
The average has been performed over
a large number of trajectories starting from random initial conditions
uniformly distributed in cell $[0,1]$ and the results are shown in
Fig.~\ref{fig:diffus} for three different perturbation amplitudes
$\varepsilon$. In all performed simultations we present in this section 
we set $\alpha=1$; the influence of $\alpha$ on diffusive behaviour 
is discussed in the following section.  

We see that diffusion is absent for
$\varepsilon$ small enough, thus, a transition between
a diffusive to non-diffusive regime is expected upon decreasing
$\varepsilon$.
When the limit in expression~(\ref{eq:diff}) exists, coefficient $D$
can be  expressed
in terms of the velocity correlation function as in Green-Kubo
formula~\cite{Dorfman,Klag}
\begin{equation}
D = \frac{1}{2}
\langle (v^0)^2 \rangle + \sum_{k=1}^{\infty} \langle v^0 v^k \rangle
\end{equation}
where $x^{i+1}-x^{i} = v^i$ can be formally regarded as a
discrete-time velocity. The simplest approximation for $D$ amounts to
assuming that velocity correlations decay so fast that only the
first term of the above series can be retained.  In the pure chaotic
system (\ref{eq:chaos}), this condition is often verified, and with
the further assumption of uniform distribution for the quantity
$x^t-[x^t]$, one obtains the approximated expression
\begin{equation}
D(a) = \frac{a}{2(1+a)}
\label{eq:Dapprox}
\end{equation}
that works rather well in the parameter region we explored \cite{Klag}.
In the non-chaotic system~(\ref{eq:nochaos}) this approximation is expected
to hold only in the large driving regime limit, where the stair-wise
structure of $F_{\Delta}$ is hidden by the perturbation effects.

The role of the external forcing on the activation of the diffusion process
in the non chaotic system has been studied by looking at the behavior
of the diffusion coefficient $D$ upon changing the perturbation
amplitude $\varepsilon$ and the ``discretization level'' defined by
the number of the micro-intervals $N$ in each cell.

\begin{figure}
\includegraphics[scale=.35]{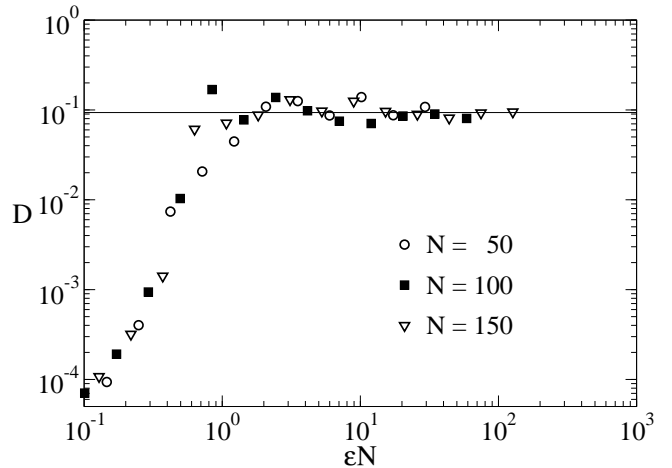}
\caption{Log-Log plot of the dependence of the
diffusion coefficient $D$ on the external forcing strength
$\varepsilon$.
Different data relative to a number of cell
micro-intervals $N=50$, $100$ and $150$ are plotted vs the natural
scaling variable $\varepsilon N$ to obtain a collapse of the curves.
Horizontal line represents the theoretical result~(\ref{eq:Dapprox})
holding for the chaotic system~(\ref{eq:chaos},\ref{eq:chaos2})}
\label{fig:thres}.
\end{figure}
Results of Fig.~\ref{fig:thres} show that $D$ is significantly
different from zero only for values $\varepsilon>\varepsilon_c$, for
which the system performs normal diffusion. Below this threshold the
values of $D$, when non zero, are so small that we cannot properly
speak of diffusion.  When $\varepsilon$ increases well above
$\varepsilon_c$, $D$ exhibits a saturation toward the value predicted
by formula~(\ref{eq:Dapprox}) (horizontal line) for the chaotic system
defined through~(\ref{eq:chaos},\ref{eq:chaos2}).  The onset of a
diffusive regime as a threshold-like phenomenon, with respect to the
external perturbation amplitude, is expected.  It can be explained by
noticing that, due to the staircase nature of the system, the
perturbation has to exceed the typical discontinuity of $F_{\Delta}$,
to activate the local instability which is the first step toward the
diffusion. Data collapsing in Figs~\ref{fig:thres} confirms this view,
because it is achieved by plotting $D$ versus the scaling variable
$\varepsilon N$. This means that, for $N$ intervals on each cell, the
typical discontinuity in the staircase-map is $O(1/N)$, then the
$\varepsilon$-threshold is $\varepsilon_c \sim 1/N$.  This behavior
is robust and does not depend on the precise shape of the forcing, as
we have verified by considering other kind of external perturbations.
However, as will be discussed in the next section, there is some
connection among the diffusive properties of the system, the periodic
orbits, the parameter $N$, and the value of the perturbation frequency
$\alpha$.

The activation of the instability performed by the perturbation
can be studied by considering a well localized ensemble of initial
conditions. We choose uniform initial distribution of ${\cal N}$ walkers 
with
$\sigma_0 ^2  \sim 10^{-12},10^{-8},10^{-6}$ and monitored the
distribution spreading under the dynamics by following the
evolution of its variance
\begin{equation}
\sigma^2(t) = \frac{1}{{\cal N}} \sum_{i=1}^{{\cal N}}
\langle (x^t_{i} - \langle x^t_{i} \rangle)^2 \rangle
\label{eq:variance}
\end{equation}
\begin{figure}
\includegraphics[scale=.35]{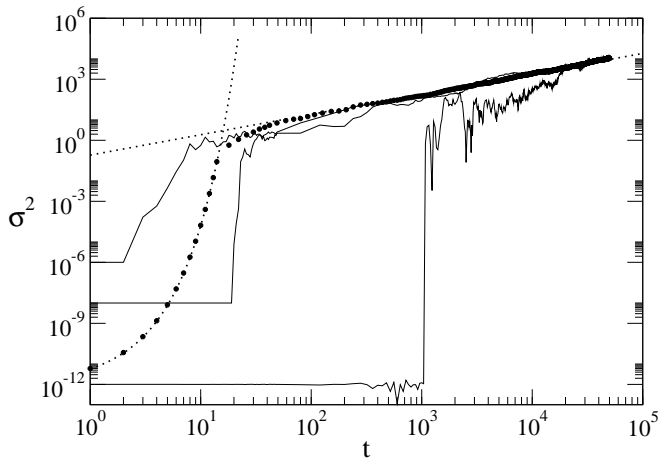}
\caption{\label{fig:finite}
Time behavior of the variance (Eq.~\ref{eq:variance})
for chaotic and non chaotic system with $a=0.23$. The discretization
level is $N=100$ for the latter.
Full lines refer to the non chaotic model and from below correspond to
initial values $\sigma_0^2=10^{-12},10^{-8},10^{-6}$ respectively.
For a direct comparison, full dots represent
$\sigma^2(t)$ in the chaotic case, with $\sigma_0^2=10^{-12}$,
which clearly exhibits a crossover from the exponential growth at small
times (governed by Lyapunov exponent) to normal diffusion at large times.
Dotted lines indicate the predicted chaotic growth $\exp(2\lambda t)$ and
diffusion $2 D(a)t$ (see formula~\ref{eq:Dapprox}).}
\end{figure}
Note that, in contrast with the average in Eq.~(\ref{eq:diff}) now the
average is over an ensemble of initial conditions, uniformly
distributed on a small size interval.  The spreading of the initial
cloud of points is better understood when we make a comparison between
the chaotic (\ref{eq:chaos}) and non chaotic system
(\ref{eq:nochaos}).  In the former, the instability makes the variance
growing exponentially at the rate $\ln[2(1+a)]$ which is its Lyapunov
exponent.  This exponential separation lasts until the typical linear
behavior of the diffusion takes place (see Fig.~\ref{fig:finite}).  In
the non-chaotic system, instead, the behavior of $\sigma^2(t)$
strongly depends on its initial value $\sigma_0^2$ and we have to
consider two sub-cases, whether $\sigma_0 \agt 1/N$ or $\sigma_0 \ll
1/N$.  When $\sigma_0 \agt 1/N$, basically one observes the same
behavior of the continuous limit i.e. $\sigma^2(t) \sim \sigma_0^2
\exp(2\lambda t)$ for times $t < t_* \sim \ln (1/\sigma_0)$ and
$\sigma^2 (t) \sim t$ for $t>t_*$.  Instead, if $\sigma_0 \ll 1/N$ we
can distinguish three different regimes: a) $\sigma (t)$ remains
constant for a certain time span $[0,t_1]$, b) the instability starts
being active and the $\sigma$ grows exponentially as in the chaotic
case, $\sigma (t) \sim (1/N) \exp[\lambda (t-t_1)]$, c) the system
eventually reaches the regime of standard diffusion where $\sigma^2$
behaves linearly.  The crossover time $t_1$ between the regimes a) and
b) depends on the size of the discontinuities of $F_{\Delta}$ and the
specific realization of the randomness and, of course, it decreases
when either $N$ or $\sigma_0$ increases.  We stress that the above
scenario is due to the presence of finite size instabilities that in a
non chaotic Lorentz gas would correspond to the defocusing mechanism
of a beam by the vertices of polygonal obstacles.

\begin{figure}
\includegraphics[scale=.35]{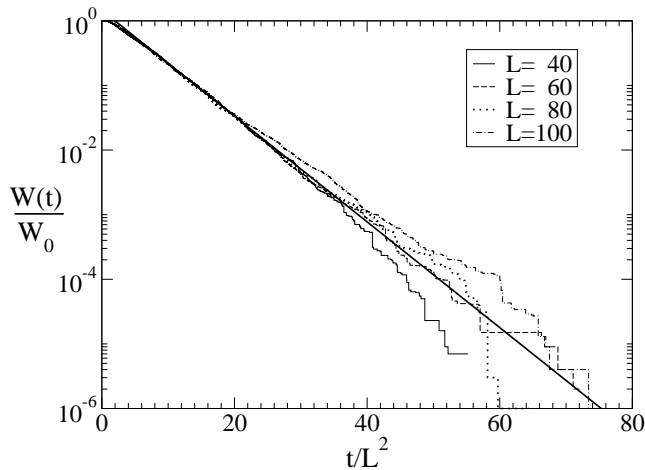}
\caption{\label{fig:escape} Exponential decay with time of the
fraction $W(t)/W_0$ of walkers that, at time $t$, never left the
region of length $L=40,60,80,100$, centered around the origin of the
lattice of cells. Model parameters are $a=0.23$, $\varepsilon=0.1$ and
$N=100$. The disorder configuration is replicated each $M=20$ cells.
The walker are initially ($t=0$) placed inside the cell $[0,1]$}
\end{figure}

The onset of standard diffusive behavior in model~(\ref{eq:nochaos})
depends on large-time features of the deterministic dynamics.  Indeed
the presence of periodic trajectories, standing or running
\cite{note}, have a strong influence on the diffusion, because the
formers tend to suppress diffusion while the latter induce a
``trivial'' ballistic behavior.  In Ref.~\cite{DC}, it has been
conjectured that the main difference between a fully chaotic
deterministic system and a non chaotic system exhibiting diffusion, as
a consequence of quenched randomness, is in the different effects the
periodic trajectories have on the systems.  An escape rate method
\cite{Dorfman,Gaspard} can be applied to analyze these effects. This
method consists on starting with an ensemble of $W(0)$ trajectories
initially localized in the region ${\cal R}_L$ between $-L/2$ and
$L/2$ and computing the number $W(t)$ of them which never left ${\cal
R}_L$ until the time $t$. For a pure random walk process one would
expect an exponential decay
\begin{equation}
W(t) \sim \exp(-\gamma t)
\label{eq:dacay}
\end{equation}
where the rate $\gamma$ is proportional to the diffusion coefficient
$D$ through the relation $\gamma \propto D/L^2$. In this sense the method
is an alternative way to estimate the diffusion coefficient from
measurement of $\gamma$. However possible deviations from
(\ref{eq:dacay}) generally indicate a non diffusive behavior related
to the presence of periodic orbits.
As clearly seen in fig~\ref{fig:escape},
on all the range of sizes $L$ we have considered,
we do not observe any substantial deviation
from the exponential decay that can be associated to periodic orbit
effects.
One can wonder whether the diffusive behavior could be merely due to
a possible non zero spatial entropy per unit length of the quenched
randomness.  To unravel the role of static randomness as possible
source of entropy and thus diffusion, following Ref.~\cite{DC}, we
have considered a system were the same configuration of the disorder
is repeated every $M$ cells ({\em i.e.}
$\xi^{(\ell)}_n = \xi^{(\ell+M)}_n$), so the
entropy for unit length is surely
zero. 
This point is crucial, because, someone can argue that a 
deterministic infinite system 
with spatial randomness can be interpreted as an effective stochastic system,
but this is a ``matter of taste''. Anyway, our system with a spatially 
periodic randonness is deterministic from any point of view.

Looking at the diffusion of an ensemble of walkers we obtain
that:
\begin{description}
\item[a)] there is weak average drift $V$, that vanishes approximately
           as  $V\sim1/M$  when $M$ goes to infinity.
\item[b)] there is still a diffusive behavior i.e. $\langle (x^t - V t)^2
\rangle \sim 2Dt$, with a coefficient $D$ very close to the value obtained
by formula~(\ref{eq:Dapprox}).
\end{description}
Let us note that this behavior is obtained by
considering single realizations of the randomness: the value of the
drift changes with the realization, in such a way that an average on
the disorder would give a null drift. The fact that $V\sim 1/M$ can be 
easily interpreted as a self-averaging property, for
large $M$. On the other hand if we write 
$\langle (x^t)^2 \rangle = (V t)^2 + 2D t$, we can define a crossover time 
$\tau_c \sim D M^2$, from pure diffusive to a ballistic regime. 
When $M$ is finite but large enough the crossover time becomes very large.   
Therefore, we can safely conclude that, even with a
vanishing (spatial) entropy density of randomness, at least for
sufficiently large $M$, the system is able to effectively perform
standard diffusion for a very long time.
However we stress that looking at $\langle (x^t - V t)^2\rangle$ we do not
observe any crossover.

\section{Space-time billiards}
As mentioned previously, in the one-dimensional maps proposed in this
paper, time plays the role of the second dimension in the
two-dimensional billiard models (e.g., Lorentz and Ehrenfest models).
Here we show this explicitly by describing how the one-dimensional
time-dependent maps define billiards in space-time.  For convenience,
we carry out the construction for model (\ref{model_2}) where the
partition of the unit cells is uniform. However, the same construction
can be done for model (\ref{model_1}).

\begin{figure}
\includegraphics[scale=.35]{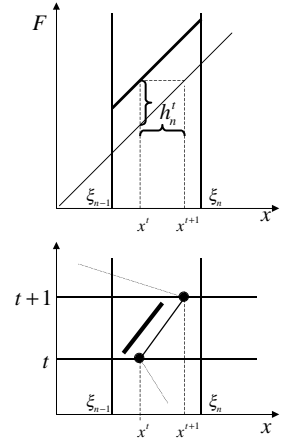}
\caption{\label{fig:spacetime}
Construction of space-time billiard from a piece-wise linear with 
unit slope map.}
\end{figure}

\begin{figure}
\includegraphics[scale=.6]{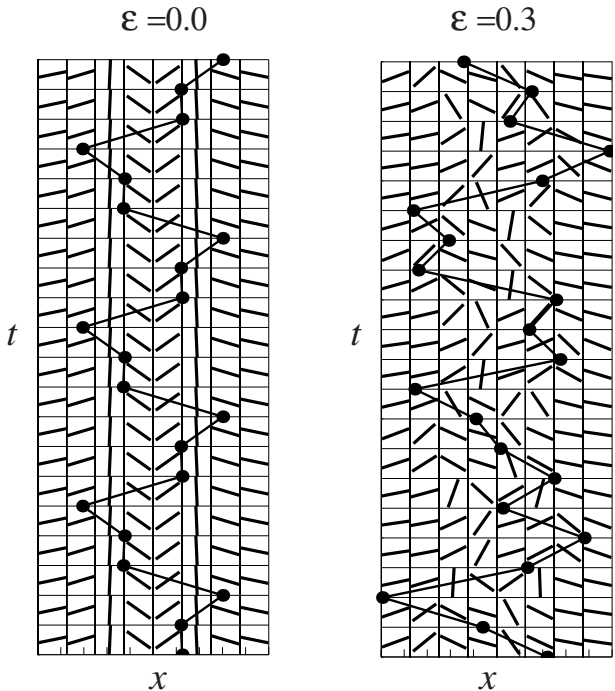}
\caption{\label{fig:storbits} Space-time billiards without time
dependence ($\varepsilon =0$), and with time dependence ($\varepsilon
\neq 0$).}
\end{figure}

Let $[\xi_{n-1}, \xi_{n}]$ be a micro-interval of a given unit cell. 
Then, for any initial
condition $x_0 \in [\xi_{n-1}, \xi_{n}]$, the velocity of the particle
at time $t$ is
\begin{equation}
\label{eq_d_2}
h_n^t=x^{t+1}-x^t=\lambda_n+ \varepsilon\, \cos (\alpha t) \, ,
\end{equation}
where $h_n^t$ is the height (including the time-dependent
perturbation) of the piece-wise linear segment defining the map
according to Eq.~(\ref{model_2}). As shown in
Fig.~\ref{fig:spacetime}, in a space-time diagram the ``world-line" of
this orbit is the straight line joining the points $(x^t, t)=(x^{t+1},
t+1)$ with slope $h_n^t$.  Accordingly, as shown in
Fig.~\ref{fig:storbits}, the space-time can be decomposed into
discrete boxes $[\xi_{n-1}, \xi_{n}] \times [t, t+1]$, and the
world-line of successive iterates of the map appears as a zig-zag
trajectory joining different boxes.  The rule of the billiard is that
a particle entering box $n$ will leave this box with speed
$h^t_n$. That is, in going from box $m$ to box $n$ a particle suffers
a space-time scattering event $\delta v= h^{t+1}_n-h^t_m$, where $\delta
v=v^{t+1}-v^t$ is the change in the particle velocity.  In the
space-time diagram the scattering rule is represented by drawing at
each box a line segment with slope given by $h^t_n$ that according to
(\ref{eq_d_2}) has a random component $\lambda_n$, and a time-periodic
component.
\begin{figure}
\includegraphics[scale=.5]{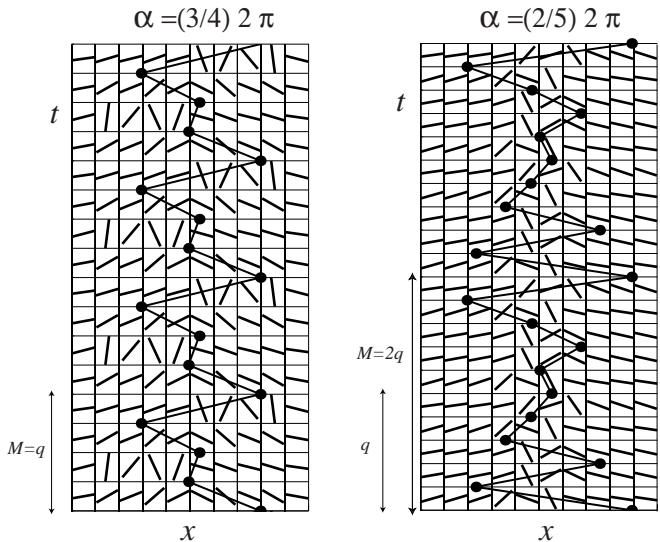}
\caption{\label{fig:stperiodic} 
Periodic orbits in a space-time
billiard with rational perturbation frequency}. 
\end{figure}
According to this, the time-dependent, zero-Lyapunov exponent (slope
one) map is equivalent to a space-time billiard with scattering rule
$h^t_n$. Note that these models are simpler than the usual 2-$d$
billiards ({\em e.g.} Lorentz gas) because the scattering rule is 
independent of the angle of incidence. 
Although this is just a reformulation of the same model, the
billiard construction provides useful insights into the problem and
allows the connection with the two-dimensional billiard problems.
When the map is time-independent ($\varepsilon=0$) the scattering rule
depends only on $n$ ($h_n^p=h_n^q$ for any integers $p$ and $q$) and
as shown in Fig.~\ref{fig:storbits}-(a) the billiard structure is
simply a copy of the billiard at $t=0$. In this case, despite the
spatial randomness, the lack of a time dependence leads to a highly
structured space-time billiard in which periodic orbits are
ubiquitous.  As a consequence, in this case there is no diffusion, a
result already mentioned before.  On the other hand, as
Figure~\ref{fig:storbits}-(b) shows, when $\varepsilon\neq 0$ the
billiard exhibits space-time disorder which leads to the possibility of
diffusive behavior.

\begin{figure}
\includegraphics[scale=.5]{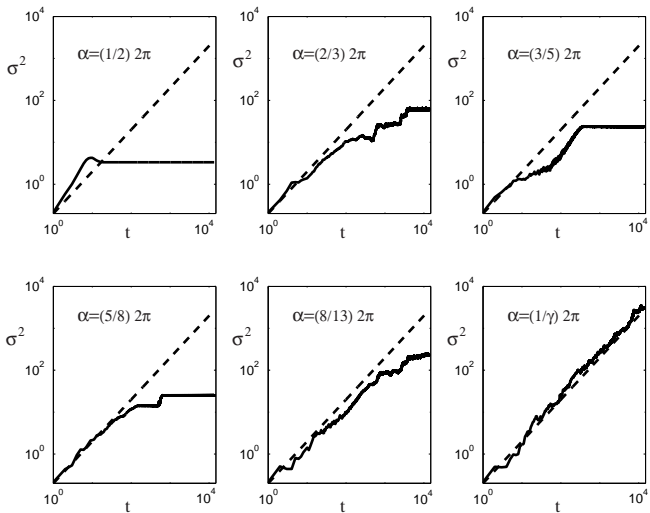}
\caption{\label{fig:gmdiffusion}
Mean square displacement as function of time
in model (\ref{model_2}) for a sequence of values of the
perturbation frequency
$\alpha$ approaching the inverse golden mean $1/\gamma$.
The heights of map are uniformely distributed in $[0,1]$ 
($\delta=0$ see text).}
\end{figure}

As discussed in Ref.~\cite{DC} for polygonal two-dimensional
billiards, periodic orbits play an important role.  The existence of
periodic orbits in the space-time billiard is intimately connected
with the frequency $\alpha$ of the time periodic perturbation. If
$\{x^t\}$ is a periodic orbit of period $M$ ($x^{t+M}=x^t$ for any
$t$), the billiard must be $q$-periodic, that is $h_n^{t+q}=h_n^t$ for
any $t$, where $M=m q$ and $m$ is a positive integer. According to
Eq.~(\ref{eq_d_2}) this will be the case provided
\begin{equation}
\label{eq_d_3}
\alpha=2\,\pi\, \nu\, ,
\end{equation}
with $\nu=p/q$.  Figure~\ref{fig:stperiodic} show two examples of
periodic orbits with $\varepsilon=0.2$.  In panel (a), $\alpha=2 \pi
(3/4)$, and the period of the billiard is the same as the period of
the orbit, $q=M=4$.  In panel (b), $\alpha=2 \pi (2/5)$, and the
period of the orbit is twice period of the billiard $M=2 q=10$.  Thus,
for rational $\nu$ the billiard shows a periodic structure in time,
and periodic orbits are likely to exist leading to the suppression of
diffusion. On the other hand, when $\nu$ is an irrational number the
billiard structure never repeats exactly in time and strictly speaking
there are no periodic orbits.  Motivated by this, it is possible to
related the degree of temporal disorder of the billiard to the degree
of irrationality of $\nu$ as determined, for example, by the continued
fraction approximation of $\nu$. In this regard, billiards with
irrational frequencies $\nu$ which are easily approximated by
rationals are ``less disordered" than billiards with hard to
approximate frequencies. To explore this ideas
Fig.~\ref{fig:gmdiffusion} shows the time evolution of the variance
for different values of the perturbation frequency in the same random
realization of model (\ref{model_2}) with $N=40$ and $\varepsilon=0.4$.
For the perturbation frequencies, $\alpha=(p/q) 2 \pi$, we took the
continued fraction expansion of the inverse golden mean
\begin{equation}
\nu=1/2,\, 2/3,\, 3/5,\, \ldots F_{n-1}/F_n\, \ldots \rightarrow
\left(\sqrt{5}-1\right)/2 = \gamma^{-1},
\end{equation}
where $F_0=F_1=1$, and $F_n=F_{n-1}+F_{n-2}$.  The chose of the golden
mean is motivated by the fact that this number is the hardest to
approximate with rational.
In the model the disorder configuration is repeated every $M$ cells, and for
small $M$ the recurrence time of an orbit (i.e. the time to return to the same
microinterval, module $M$) decreases. Since we are interested
in the role of periodic orbits, here we consider the limit case $M=1$. That is,
in the simulations reported in Fig.~\ref{fig:gmdiffusion}, the disorder
configuration in the unit
interval $[0,1]$ was copied to all the cells $[\ell, \ell+1]$.
As expected the figure shows the
suppression of diffusion for rational frequencies. However, more
interesting is the fact that there is a tendency for the onset of the
suppression of diffusion to increase as $\nu$ approaches
$1/\gamma$. This is related to the fact that for rationals with lower
denominators the time period of the billiard is small, and the
probability that the recurrence time of an orbit
coincides with a
multiple of the period of the billiard is high.  The recurrence time
is also related to the value of $N$. When $N$ is large, the
probability for an initial condition to return to the same
micro-interval (module $M$) is small, and  periodic orbits are more
scarce. This explains why for large enough $N$ and $M$, diffusion might not be
suppressed even for rational frequencies with large denominators.
This suggests that, for large enough $N$ and $M$, deviations from diffusion
are unlikely to be observed numerically in the case of rational 
frequencies with large denominators.  

\section{Conclusions and remarks}
From the above results we can conclude
that two basic ingredients are relevant for diffusion,
1) a finite size instability mechanism: in the chaotic system this is
given by the Lyapunov
instability, in the "stepwise" system this effect stems from the jumps
2) a mechanism to suppress periodic orbits and therefore to allow for a
diffusion at large scales.
In the presence of ``strong chaos''
(i.e. when all the periodic orbits are unstable) point 2) is automatically
guaranteed thus the periodic orbits formalism \cite{Artuso,CEG95,
Dorfman} can be fully applied to compute
the diffusion coefficient. In non chaotic systems, however, like the present
one, the mechanism 2) is the result of a combined effect of quenched randomness
and quasi-periodic forcing.
 
In summary, we have introduced and studied models exhibiting diffusion
in the absence of any source of chaotic behavior. The models
represent the $1$-$d$ analogue of non chaotic Lorentz gas (i.e. with
polygonal obstacles) discussed by other authors in connection with the
debate around experimental evidences for a distinction between chaotic
and stochastic diffusion.
Our results indicate that when the chaotic instability condition (positive 
Lyapunov exponent) is replaced by the presence of finite size
instability (non positive Lyapunov exponent) we need both an external
quasi-periodic perturbation and disorder for preventing the system falling
into either a periodic/quasi-periodic evolution or a drift dominated
behavior.

We believe that this kind of behaviour is rather general, in the sense
that chaos does not seem to be a necessary condition for the validity
of some statistical features. This is in agreement with the recent results
of Ref.~\cite{LRB}, where the applicability of the generalized Gallavotti-Cohen
fluctuation formula \cite{GC} has been proven for non chaotic systems too. 

\section{Acknowledgments}
This work has been partially supported by MIUR (Cofin.
{\em Fisica Statistica di Sistemi Classici e Quantistici}).
D dCN was supported by the Oak Ridge National Laboratory, managed by 
UT-Battelle, LLC,
for the U.S. Department of Energy under Contract No. 
DE-AC05-00OR22725.  We thank
Massimo Cencini and Stefano Ruffo for useful suggestions and discussions. D dCN
gratefully acknowledges  the hospitality of the Department of Physics 
of the University of
Rome ``La Sapienza" during the elaboration of this work.



\begin{references}
\bibitem{note} A periodic trajectory ${x_t}$, of period $p$, is called
                standing when we have $x^{t+p} = x^{t}$,
                while is called running when $x^{t+p} = x^{t} + n_p$,
                $n_p$ being an  integer indicating the lattice traslation.
\end{references}
\end{document}